\documentclass{desyproc}

\begin{document}
\title{Non-minimally coupled Scalar Dark Matter from Inflationary Fluctuations}

\author{{\slshape Gonzalo Alonso-\'Alvarez$^1$}\\[1ex]
$^1$Institute for Theoretical Physics, Heidelberg University, Heidelberg, Germany}

\contribID{Lindner\_Axel}

\confID{20012}  
\desyproc{DESY-PROC-2018-03}
\acronym{Patras 2018} 
\doi  

\maketitle

\begin{abstract}
It is well known that light scalar fields present during inflation are coherently excited. We show that if the field couples to gravity in a non-minimal way, the fluctuations at large scales are suppressed with respect to the small scales ones. This fact allows for the field excitations to make a sizeable contribution to the energy density of the universe without generating too large isocurvature fluctuations at observable scales. We show that this mechanism could generate all the observed dark matter and study the main cosmological implications of this setup.
\end{abstract}

\section{Introduction}
\emph{The material presented here is based on \cite{Alonso-Alvarez:2018tus}, to which we refer for more in-depth explanations and discussion, together with a more complete set of references.}

Gravitational observations in cosmological and astrophysical settings strongly suggest that a large fraction of the energy density of our Universe is stored in some form of dark matter.
New, weakly interacting and massive bosonic degrees of freedom are well motivated ways to explain this.
Masses in a wide range of scales are viable, from the ultra light fuzzy dark matter limit to very heavy fields lying close to the Planck scale.
Different observables and experimental techniques have been developed to explore this paradigm.
The GeV-TeV range has been the most extensively tested, but new ideas will allow to probe lighter fields down to the meV in the future.
Some astrophysical observations give a hint for dark matter with a mass around the keV, further motivating theoretical studies of this intermediate mass range.

The two standard frameworks of bosonic dark matter production, thermal freeze out and the misalignment mechanism, require modifications or tuning of parameters to yield the correct relic abundance for these intermediate mass values.
It is therefore timely to explore alternative mechanisms that could explain the origin of dark matter in this mass range.

It was recently shown in \cite{Graham:2015rva} that dark photon dark matter can be generated from fluctuations of inflationary origin.
We argue that an analogue production occurs for a scalar field non-minimally coupled to gravity.
A similar idea was used in \cite{Cosme:2018nly} in the context of Higgs portal dark matter.
The non-minimal coupling suppresses the amplitude of the fluctuations at large scales, which explains why isocurvature perturbations have not been observed by the Planck mission~\cite{Akrami:2018odb} in the cosmic microwave background (CMB).
We study the quantum generation of the fluctuations and their cosmological evolution to obtain the late-times power spectrum.
This allows us to make precise quantitative statements regarding the present abundance of dark matter and the particular features of the mechanism.

\section{Non-minimal coupling to gravity}
The action for a massive scalar field non-minimally coupled to gravity that we consider is
\begin{equation}\label{eq:action}
S = \int \mathop{\mathrm{d}^4x} \sqrt{-g} \left( \frac{1}{2} \left( m_{\rm Pl}^2 - \xi \phi^2\right) R - \frac{1}{2} \partial_\mu \phi \partial^\mu \phi - \frac{1}{2} m^2\phi^2 \right).
\end{equation}
We use the mostly plus convention and focus on positive values of the coupling constant $\xi$.
The usual Einstein equations should be modified to take into account the presente of the direct coupling between $\phi$ and the Ricci scalar $R$.
The modifications induced are however suppressed by powers of $\phi/m_{\rm Pl}$ and remain small as long as $\phi$ doesn't probe field values close to the Planck scale.
This condition is amply satisfied in our setup and we can consequently neglect any backreaction effects on the geometry.

We specify to a Florian-Lemaitre-Robertson-Walker (FLRW) background.
The Ricci scalar is given by $R=3(1-3\omega)H^2$, $H$ being the Hubble parameter and $\omega=p/\rho$ the equation of state of the fluid driving the expansion.
The classical equation of motion for $\phi$ reads
\begin{equation}\label{eq:classical_eom_k}
\left(\partial^2_t + 3H\partial_t + \frac{k^2}{a^2} + m^2 + \xi R \right)\phi(k,t) = 0.
\end{equation}

\section{Cosmological evolution}
\subsection{Quantum evolution during inflation}
Let us first argue that any preinflationary homogeneous initial condition $\phi_0$ is washed out due to the non-minimal coupling.
We solve the $k\rightarrow 0$ limit of Eq.~\eqref{eq:classical_eom_k} adopting a constant value of the Hubble parameter during inflation.
The result is an exponential suppression,
\begin{equation}
\phi_\mathrm{E} \simeq \phi_0\ \mathrm{e}^{-\frac{\mathrm{Re}(\alpha_-)}{2}N},
\end{equation}
where $\alpha_- = 3-\sqrt{9-48\xi}$ and $N$ is the total number of efolds of inflation.
However, even if the average field $\left<\phi\right>$ is quickly damped away, the two-point correlator $\left<\phi^2\right>$ receives sizeable contributions from non-vanishing momentum modes due to the growth of quantum fluctuations.

This effect can be understood within the framework of quantum field theory in classically curved backgrounds.
A computation leads to the power spectrum of a mode of comoving momentum $k$:
\begin{equation}\label{eq:inflationary_power_spectrum}
\mathcal{P}_\phi(k,a(t)) = \left( \frac{H_I}{2\pi} \right)^2 \frac{\pi}{2}\, \left( \frac{k}{a(t) H_I} \right)^3 \left| H_\nu^{(1)} \left( \frac{k}{a(t) H_I} \right) \right|^2.
\end{equation}
Here, $H_\nu^{(1)}$ is the Hankel function of the first kind and $\nu$ is given by $\nu^2 = 9/4 - 12\xi - m^2/H_I^2$.
This power spectrum presents a divergence for large momenta (it grows as $k^2$), which means that the variance of the field seems to be dominated by the ultraviolet modes.
We find a regularization scheme that provides an unambiguous and finite value of the energy density at late times.
Being assured that this procedure leads to well-defined observables in the late universe, we can evaluate the regularized version of Eq.~\eqref{eq:inflationary_power_spectrum} for modes that are well outside the horizon, i.e. at a time when they satisfy $k\ll a(t)H_I$.
\begin{equation}\label{eq:primordial_power_spectrum}
\mathcal{P}_\phi^{\mathrm{(reg)}} (k,t) \simeq \mathcal{P}_{\phi_0}^{\mathrm{(reg)}}(k)\ \left( \frac{k}{a(t)H_\mathrm{I}} \right)^{3-2\nu}, \quad\mathrm{with}\quad
\mathcal{P}_{\phi_0}^{\mathrm{(reg)}}(k) = \left( \frac{H_\mathrm{I}}{2\pi} \right)^2 \frac{2^{2\nu-1}\Gamma^2(\nu)}{\pi}.
\end{equation}
This result serves as the initial condition for the posterior evolution.

\begin{figure}[t]
\centerline{\includegraphics[width=0.6\textwidth]{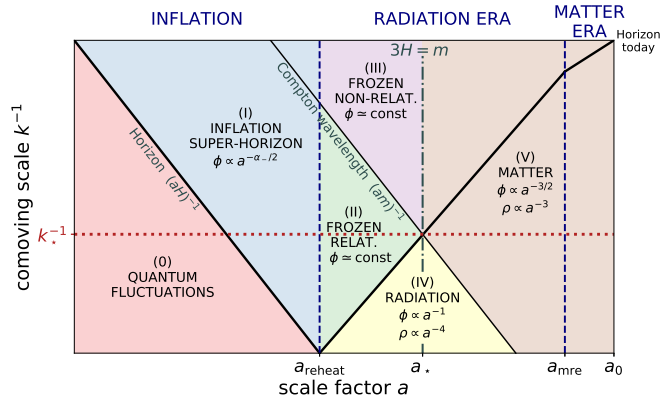}}
\caption{(From \cite{Alonso-Alvarez:2018tus}). Cosmological evolution of momentum modes as a function of their comoving wavelength $k^{-1}$.
The evolution is distinct in each regime represented by a different color, according to Eq.~\eqref{eq:classical_eom_k}.
The modes evolve in horizontal lines from left to right in the figure.
They start their life as small quantum fluctuations during inflation (0), whose amplitude grows as they exit the horizon.
In the superhorizon inflationary regime (I), the large effective mass suppresses their amplitude.
Once inflation ends, the field acquires its late time mass and becomes frozen on superhorizon scales (II and III).
If it reenters the horizon while still relativistic (IV), it oscillates with a damped amplitude.
At late times when the modes are non-relativistic and Hubble friction is overcome (V), the modes redshift like pressureless matter~\cite{Arias:2012az}.}\label{fig:master_plot}
\end{figure}

\subsection{Classical evolution and late time observables}
After horizon exit, the modes transition into a regime in which they can be treated classically.
This means that their evolution can be traced by solving the classical equation of motion Eq.~\eqref{eq:classical_eom_k}.
The power spectrum is obtained by evolving the initial condition Eq.~\eqref{eq:primordial_power_spectrum}.

The solution of the equation of motion in the different regimes of expansion of the Universe is summarized in Fig.~\ref{fig:master_plot}.
The power spectrum of the field at late times is dominated by the modes close to $k_\star$ (highlighted in red), as this is the mode that gets suppressed the least during the cosmological evolution.
The corresponding comoving scale is $k_\star^{-1} \simeq 4.1\cdot 10^{7}\ \mathrm{km}\cdot \sqrt{\mathrm{eV}/m}$, which is orders of magnitude below the ones accessible to cosmological probes.

\begin{figure}
\centering
\begin{minipage}{.46\textwidth}
  \centering
  \includegraphics[width=\linewidth]{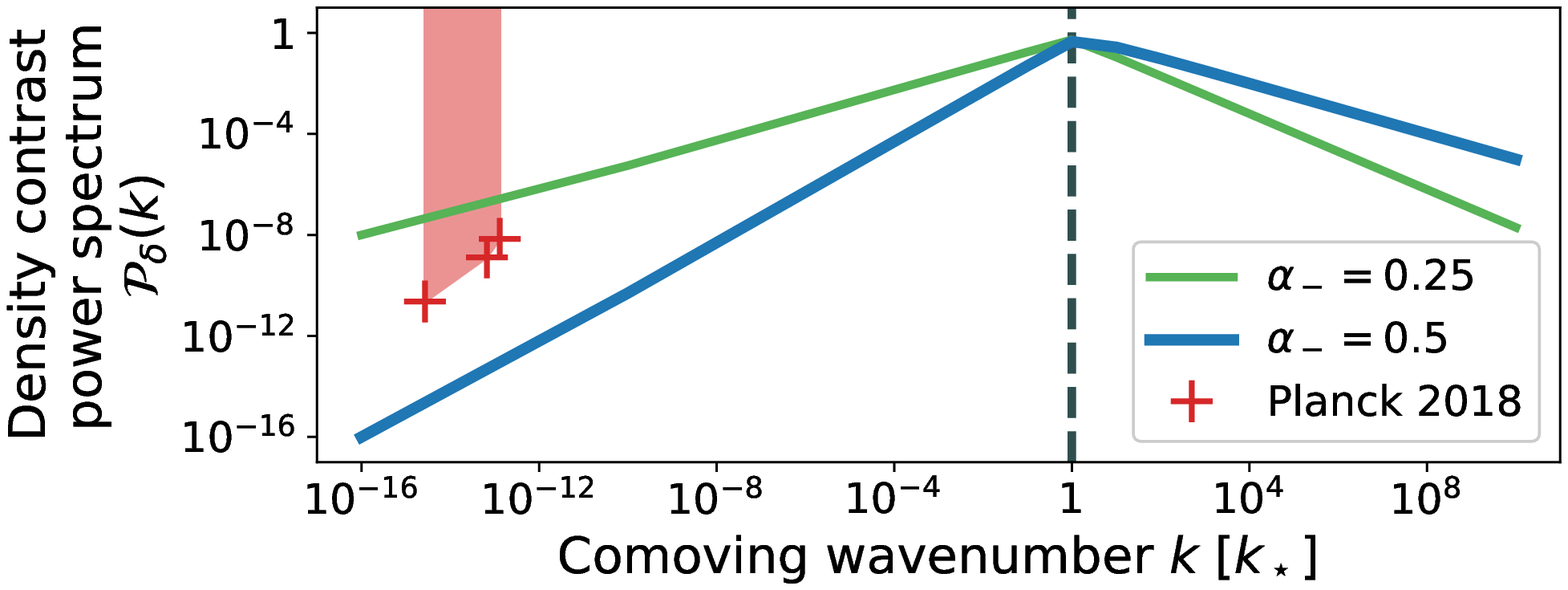}
  \caption{(From \cite{Alonso-Alvarez:2018tus}). Isocurvature power spectrum. It peaks at the scale $k_\star^{-1}$ and drops below Planck limits at large scales.}
  \label{fig:power_spectrum}
\end{minipage}\hspace{10mm}
\begin{minipage}{.46\textwidth}
  \centering
  \includegraphics[width=.7\linewidth]{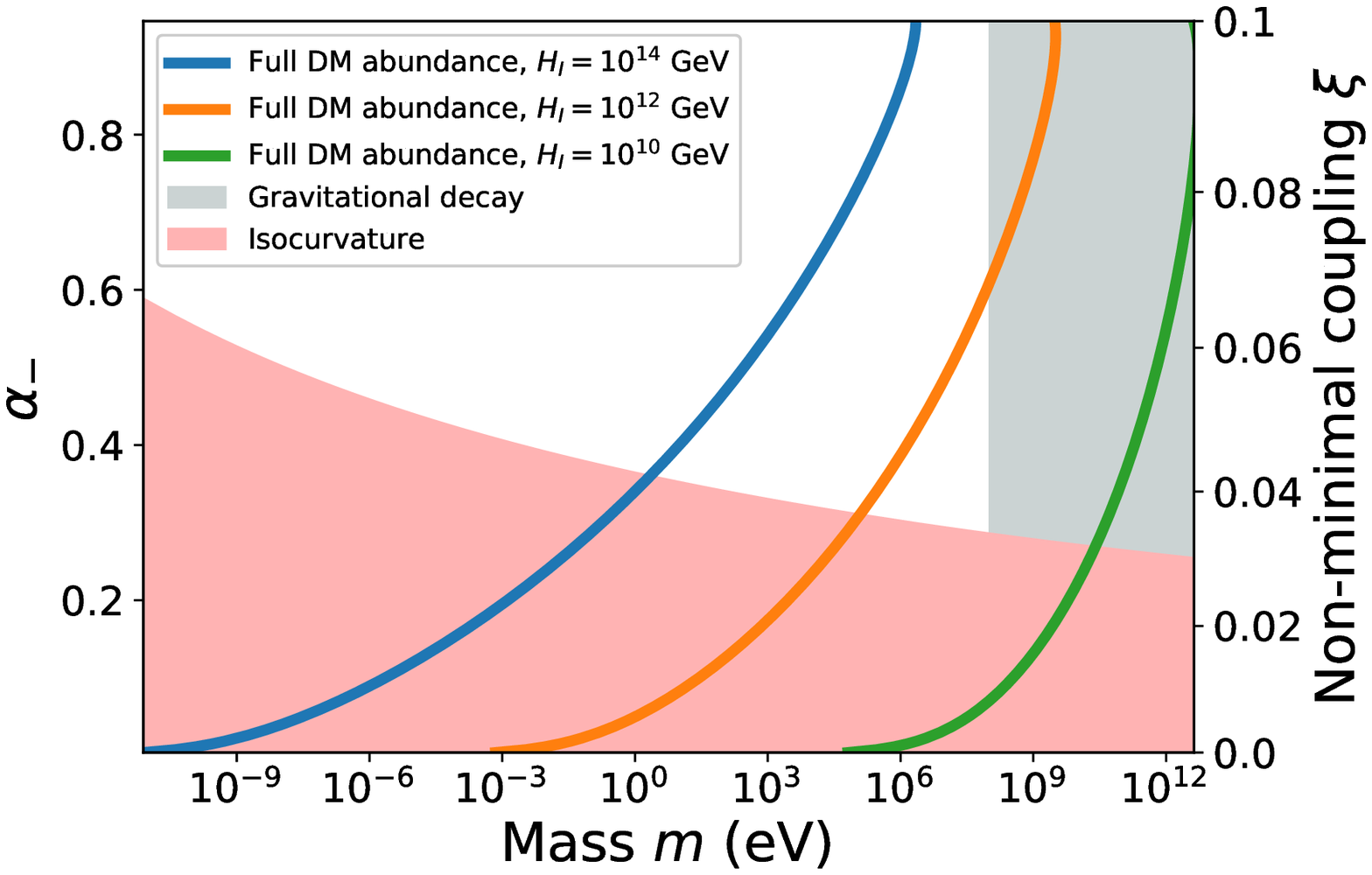}
  \caption{(From \cite{Alonso-Alvarez:2018tus}). Parameter space of the model. See text for more details.}
  \label{fig:relic_density}
\end{minipage}
\end{figure}

The amplitude of isocurvature perturbations at CMB scales is read off the density contrast power spectrum, which can be derived from the field power spectrum.
The result is shown in Fig.~\ref{fig:power_spectrum}.
For a large enough non-minimal coupling, the amplitude of isocurvature perturbations at CMB scales falls below the constraints set by Planck.
The peaked power spectrum implies that the bulk of the energy density is stored in overdensities of typical comoving size $k_\star^{-1}$.

We obtain the relic abundance of the field and compare with the observed dark matter one,
\begin{equation}\label{eq:dark_matter_abundance}
\frac{\Omega_\phi}{\Omega_{\mathrm{DM}}} \simeq C(\alpha_-) \frac{1}{m_{\rm pl}^2}\, H_{\mathrm{eq}}^{-\frac{1}{2}}\ H_\mathrm{I}^{\frac{1}{2} (4-\alpha_-)}\ m^{\frac{1}{2} (\alpha_-+1)},
\end{equation}
where $C(\alpha_-)$ is an $\mathcal{O}(1)$ factor.
The different lines in Fig.~\ref{fig:relic_density} show values of the scale of inflation for which the dark matter abundance is reproduced, as a function of the mass and the non-minimal coupling.
The red region in the figure is excluded by the Planck isocurvature limit and in the grey area graviton-mediated decay could render the field cosmologically unstable.

\section{Summary and discussion}
Generated in a purely gravitational way, the dark matter in this scenario is a (pseudo)scalar with a small non-minimal coupling to gravity.
Quantum fluctuations of the field are amplified during the inflationary epoch in the very early Universe, and the energy density thus produced behaves as non-relativistic, pressureless matter at late times.
The density perturbations are of isocurvature type, but the the non-minimal coupling suppresses their amplitude at large scales.
The scenario thus avoids the stringent constraints on dark matter isocurvature set by Planck.

As overdensities around the comoving scale $k_\star^{-1}$ are very large, we expect them to collapse and form bound structures early on.
The presence of such clumps of dark matter, together with a potentially observable level of isocurvature, are the most particular features of this setup.
Although the production mechanism presented is purely gravitational, sufficiently weak couplings to the visible sector are compatible with it.

\section{Acknowledgements}
The author would like to thank Joerg Jaeckel for his collaboration on the work presented here. This project has received funding from the European Union's Horizon 2020 research and innovation programme under the Marie Sklodowska-Curie grant agreement No 674896 (ITN ELUSIVES). The author acknowledges support by a ``La Caixa" foundation fellowship.

\section{Bibliography} 
\begin{footnotesize}
\begin{footnotesize}

\end{footnotesize}

\end{footnotesize}


\end{document}